\documentstyle[aps,floats,psfig]{revtex}

\begin{document}
\psfigurepath{.:plot:figure}
\twocolumn[\hsize\textwidth\columnwidth\hsize\csname @twocolumnfalse\endcsname
\bibliographystyle{unsrt}
\preprint{mstr3.tex, \today}
\title{
Incommensurate magnetic structure of CeRhIn$_5$
}
\author{Wei Bao, 
P.G. Pagliuso, J.L. Sarrao, 
J.D. Thompson and Z. Fisk$^{*}$}
\address{Los Alamos National Laboratory, Los Alamos, NM 87545}
\author{J. W. Lynn and R. W. Erwin}
\address{NIST Center for Neutron Research, National Institute of Standards 
and Technology, Gaithersburg, MD 20899}
\date{\today}
\maketitle
\begin{abstract}
The magnetic structure of the heavy fermion antiferromagnet
CeRhIn$_5$ is determined using neutron diffraction. We find a
magnetic wave vector ${\bf q}_M=(1/2,1/2,0.297)$, which is temperature
independent up to $T_N=3.8$~K. A staggered moment of 0.374(5)$\mu_B$ at
1.4~K, residing on the Ce ion, spirals transversely along the $c$ axis.
The nearest neighbor moments on the tetragonal basal plane are
aligned antiferromagnetically.
\end{abstract}
\vskip2pc]
%\pacs{PACS numbers: }

\narrowtext

CeRhIn$_5$ is a recently reported heavy Fermion compound that orders
antiferromagnetically at $T_N=3.8$~K\cite{hegger}.  
As a function of applied pressure,
magnetism gives way to superconductivity with $T_C=2.1$~K at 16 kbar.  
Unlike the smooth evolution of $T_N$ to zero reported for other Ce-based
heavy-Fermion antiferromagnetic superconductors, e.g., CeIn$_3$\cite{cmbr}, 
in CeRhIn$_5$ the transition is apparently first-order\cite{hegger}.  
CeRhIn$_5$ crystallizes in the tetragonal HoCoGa$_5$ structure 
($a=4.65 \AA$, $c=7.54\AA$ at 295~K),
which can be viewed, at least schematically, as alternating layers of CeIn$_3$
and RhIn$_2$\cite{nstru}.  It has been suggested that the resulting
quasi-two-dimensional configuration of the CeIn$_3$ layers is at the heart of
the unusual pressure dependence of CeRhIn$_5$.

Magnetic susceptibility  reveals a factor of two anisotropy at $T_N$ for 
$H \parallel c$ versus $H \perp c$.  
$^{115}$In NQR measurements indicate that in the N\'{e}el state the
ordered moments lie strictly in the CeIn$_3$ plane and suggest a spiral spin
structure along the $c$ axis\cite{nqr}.  The ordered moment 
develops much faster
than would be expected in mean field theory and is inferred to saturate at
about 0.1 $\mu_B$/Ce.  
Heat capacity\cite{new1} and NQR measurements\cite{nqr} also point to a
partial gapping of the Fermi surface coincident with the N\'{e}el transition
that is suggestive of a spin-density-wave (SDW) state.

The magnetic behavior of CeRhIn$_5$ should be contrasted with that of CeIn$_3$,
which orders antiferromagnetically at 10 K.  Neutron diffraction
measurements reveal an ordered moment of  0.48-0.65$\mu_B$/Ce and a
commensurate ordering wave
vector (1/2,1/2,1/2)\cite{cein,ssc}.  The development of the ordered moment 
below $T_N$
is consistent with mean field theory.  
The striking differences between CeIn$_3$ and CeRhIn$_5$ certainly
suggest that a complete determination of the magnetic structure of the
latter compound using neutron diffraction is indicated.

Single crystals of CeRhIn$_5$ were grown from an In flux as reported
previously\cite{hegger}.  The resulting crystals are well-faceted 
rectangular parallelepipeds. Neutron scattering experiments were performed
at NIST using the thermal triple-axis 
spectrometers BT9  with neutrons of energy
$E=14.7$~meV, and BT2 with neutrons
of $E=35$~meV. Pyrolytic graphite (PG) (002) was used
as the monochromator, as well as the analyzer when it was used in some 
scans.
PG filters of 3 or 5 cm thickness were used to remove higher order
neutrons. The horizontal collimations are 40-40-40-open at BT9 and
60-40-40-open at BT2.
Sample temperature was regulated by a top loading pumped He cryostat. 
Lattice parameters at 1.2-7~K are $a=4.64 \AA$ and $c=7.51\AA$.

A powder sample, ground from 10 grams of crystals and inside 
a $2.5\times 5\times 0.15$ cm Al container, was first examined 
at BT9. No peak showing
temperature dependence between 1.2 and 7~K could be detected above the
incoherent scattering background. 
This puts an upper limit for the intensity of 
any magnetic peak at 15\% of that of the (100) structural peak.

\begin{figure}[bt]
\centerline{
\psfig{file=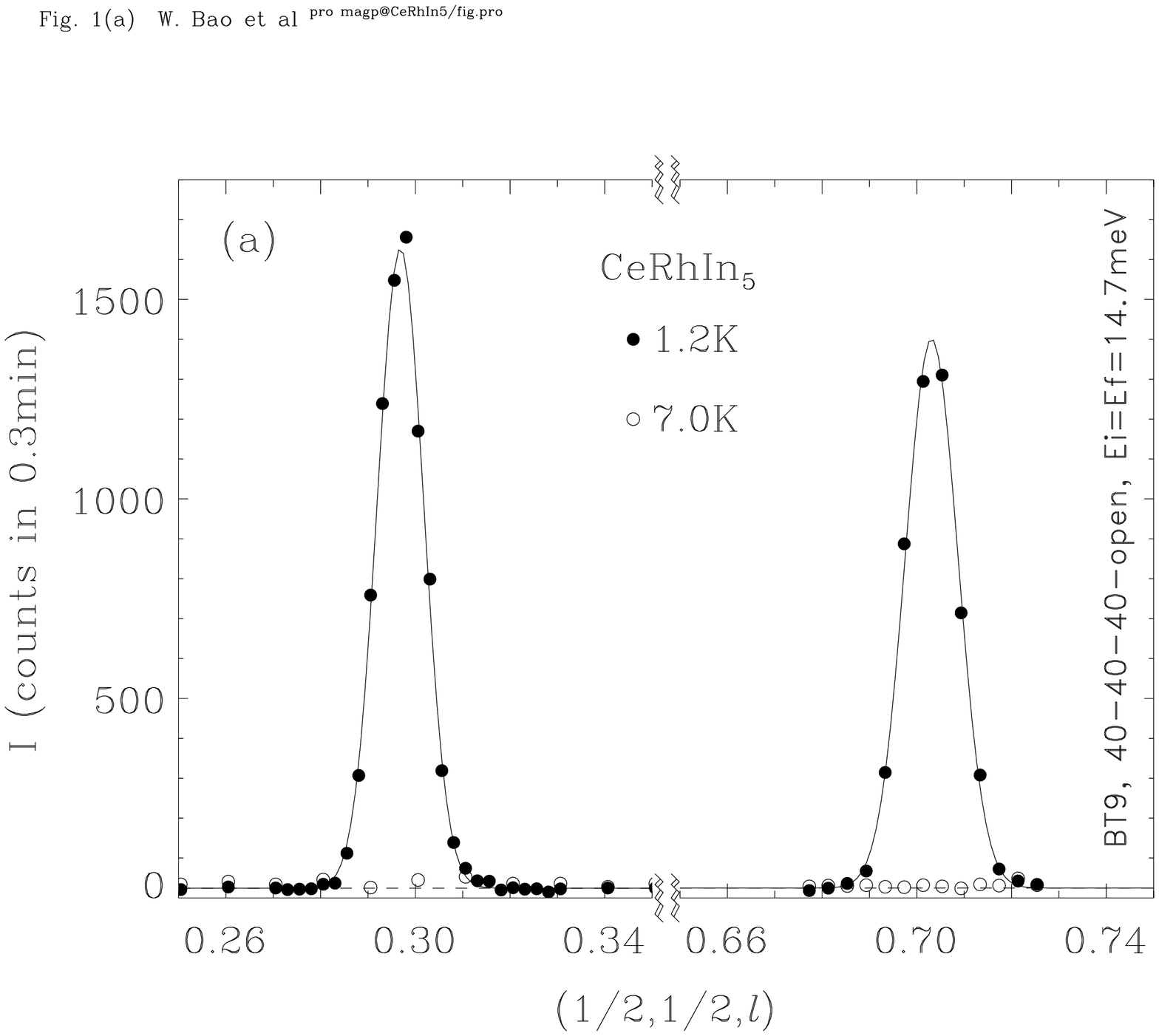,width=\columnwidth,angle=90,clip=}}
\vskip -2pc
\centerline{
\psfig{file=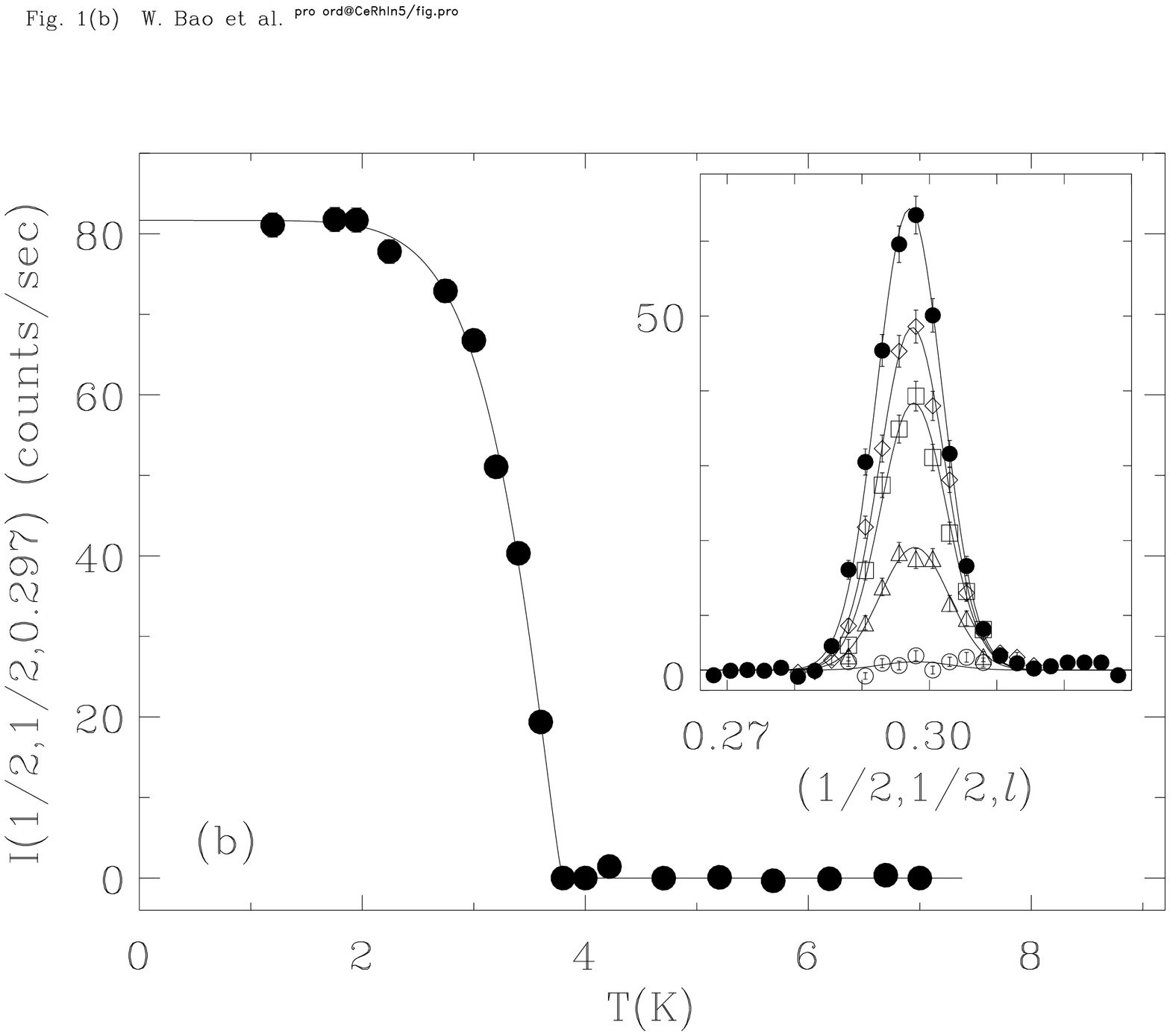,width=.9\columnwidth,angle=90,clip=}}
\caption{(a) Elastic scan through a pair of magnetic Bragg points at
1.2 and 7~K. (b) Temperature dependence of the (1/2,1/2,0.297) Bragg
peaks. Insert: Scans along the $c$ axis, with decreasing intensity,
at 3.0, 3.2, 3.4, 3.6 and 3.8~K.}
\label{magp}
\end{figure}

A roughly disk-shaped single crystal of $\sim$1 cm in diameter 
and $\sim$3 mm in thickness,
with the ``disk'' surface of the (001) plane, was used at BT9 to search for 
temperature dependent peaks. They were found as satellites of structural peaks
with an incommensurate magnetic wave vector ${\bf q}_M=(1/2,1/2,\delta)$.
No higher order harmonics can be detected at 2${\bf q}_M$ and 3${\bf q}_M$.
The upper limits for them are 2\% and 0.2\% of the $(1/2,1/2,\delta)$ peak,
respectively, at 1.6~K.
Fig.~\ref{magp}(a) shows a pair of the satellite peaks and the 
incommensurability $\delta=0.297$.

The intensity of the (1/2, 1/2, 0.297) Bragg peak
is shown in Fig.~\ref{magp}(b) as the square of
the order parameter of the magnetic phase transition. A N\'{e}el temperature
of 3.8~K is consistent with that determined from
heat capacity and resistivity\cite{hegger}. The rapid development of the
ordered moment with temperature is in agreement with the NQR
measurements\cite{nqr}. As shown by the scans at various 
temperatures in the insert to Fig.~\ref{magp}(b),
there is no detectable change in the incommensurability, $\delta$,
as a function of temperature. Additionally,
there is no detectable broadening
of the peak at this spatial resolution up to 3.6~K. In this regard,
the incommensurate magnetic structure in CeRhIn$_5$ resembles that found
in metallic V$_{2-y}$O$_3$\cite{bao93} which is due to formation of a SDW 
by electrons on the nesting part of Fermi surface. However, the 
relation between the magnetic order parameter and resistivity for 
a SDW, observed in Cr and metallic
V$_{2-y}$O$_3$\cite{bao93,Cr_dbmtmr}, seems absent in CeRhIn$_5$
(refer to Fig.~3 in \cite{hegger} and Fig.~1(b) here).

The magnetic wave vector, ${\bf q}_M = (1/2,1/2,\delta)$, already
determines that the magnetic moments of Ce or Rh ions in an $a$-$b$ plane have a 
simple nearest-neighbor antiferromagnetic arrangement which changes
incommensurately with a pitch $\delta$ along the $c$-axis. 
This incommensurate change along 
$c$, in general, for magnetic moments at $nc$, takes the form
\begin{equation}
{\bf M}=M {\rm Re}\left[({\bf x}+\alpha {\bf y})
e^{i2\pi n \delta} \right],
\end{equation}
where $\alpha$ is a complex number, {\bf x} and {\bf y} are {\em any} two
perpendicular unit vectors, and $M$ is the magnitude of the magnetic moment.
A trivial overall phase has been ignored here. To determine the
remaining variables,
a reasonable set of magnetic Bragg peaks needed to be measured.

To minimize significant absorption corrections associated
with the large disk-shaped 
crystal, a bar-shaped single crystal of cross section 
$\sim$$1.5\times 3$ mm in the scattering plane
(sides along the $c$ and (110) directions respectively)
was measured at BT2 with neutrons of $E=35$~meV. At this energy,
the neutron penetration length is 1.7~mm.
A PG filter of 5 cm thickness
is sufficient to suppress high order neutrons since an additional
PG filter did not change the intensity ratio among Bragg peaks (003), (220),
(111) and (1/2, 1/2, 0.297).
Equivalent structural Bragg peaks, such as
(111), (11$\overline{1}$) and ($\overline{1}\overline{1}\overline{1}$) 
with widely different rocking angles, have
similar intensity, indicating that the sample is mostly transparent
to the neutron beam. 

\begin{figure}[bt]
\centerline{
\psfig{file=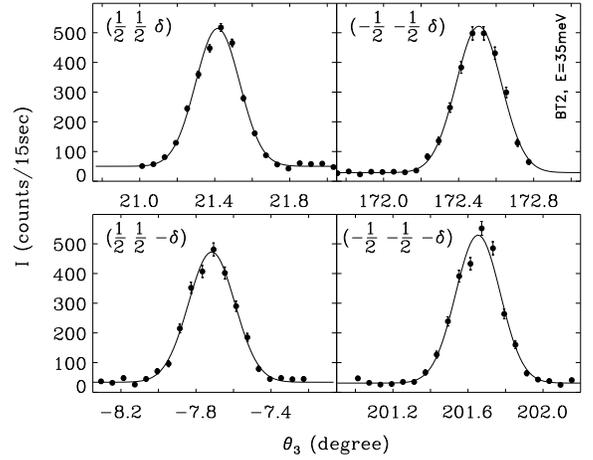,width=\columnwidth,angle=90,clip=}}
\caption{Elastic rocking scans through magnetic Bragg points 
\{1/2,1/2,0.297\} at 1.4~K.}
\label{peak4}
\end{figure}

Rocking scans for the quartet of the 
\{1/2,1/2,0.297\} magnetic Bragg peaks in Fig.~\ref{peak4} also
show similar intensity. Besides reinforcing the rocking
angle independence for Bragg peaks, the symmetry of these intensities
places useful limits on the possible magnetic moment orientation.

Integrated intensities of 20 independent magnetic Bragg peaks were measured
using rocking scans, such as those in Fig.~\ref{peak4}. 
The scattering cross-sections, 
$\sigma({\bf q})=I({\bf q}) \sin(\theta_4)$,
normalized to structural Bragg peaks
\{111\} to yield the absolute intensity, 
are listed in Table~\ref{mlist}. In such units,
\begin{table}[b]
\caption{Magnetic Bragg intensity, defined in Eq.~(\ref{eq_cs}), at
1.4~K in units of $10^{-3}$barns per CeRhIn$_5$.
}
\label{mlist}
\begin{tabular}{cd|cd}
${\bf q}$ & $\sigma$ & ${\bf q}$ & $\sigma$\\
\hline
     (0.5     0.5     0.297 ) &      8.9(2) &
     (0.5     0.5      5.297 ) &      3.7(4) \\
     (0.5     0.5     0.703 ) &      10.8(2) &
     (0.5     0.5      5.703 ) &      3.3(5) \\
     (0.5     0.5      1.297 ) &      13.2(2) &
      (0.5     0.5      6.297 ) &      2.7(5) \\
    (0.5     0.5      1.703 ) &      12.6(1) &
      ( 1.5      1.5     0.297 ) &      3.4(1) \\
    (0.5     0.5      2.297 ) &      11.0(1) &
     ( 1.5      1.5     0.703 ) &      3.6(2) \\
     (0.5     0.5      2.703 ) &      9.5(1) &
     ( 1.5      1.5      1.297 ) &      3.1(5) \\
     (0.5     0.5      3.297 ) &      7.7(2) &
     ( 1.5      1.5      1.703 ) &      3.1(5) \\
     (0.5     0.5      3.703 ) &      6.8(5) &
     ( 1.5      1.5      2.297 ) &      3.1(2) \\
     (0.5     0.5      4.297 ) &      6.3(7) &
     ( 1.5      1.5      2.703 ) &      2.9(3) \\
     (0.5     0.5      4.703 ) &      5.9(8) &
     ( 2.5      2.5     0.297 ) &      1.1(4) \\
\end{tabular}
\end{table}
\begin{equation}
\sigma({\bf q})=\left(\frac{\gamma r_0}{2}\right)^2
	\langle M\rangle^2 \left|f(q)\right|^2 
	\sum_{\mu,\nu}(\delta_{\mu\nu}
	-\widehat{\rm q}_{\mu}\widehat{\rm q}_{\nu})
	{\cal F}^*_{\mu}({\bf q}){\cal F}_{\nu}({\bf q}),
\label{eq_cs}
\end{equation}
where $(\gamma r_0/2)^2=0.07265$~barns/$\mu_B^2$, $M$ is 
the staggered moment, $f(q)$ the atomic
form factor, $\widehat{\bf q}$ the unit vector of ${\bf q}$,
and ${\cal F}_{\mu}({\bf q})$ the $\mu$th
Cartesian component of magnetic structure factor per CeRhIn$_5$.

The pattern of magnetic Bragg peaks (Table~\ref{mlist} and Fig.~\ref{peak4})
indicates that ${\bf z}\equiv
{\bf x}\times{\bf y}$ is parallel to the $c$ axis. We now consider
two cases: (I) $\alpha=\pm i$, a magnetic spiral, and 
(II) $\alpha=0$, a collinear magnetic moment modulation. For model I,
the cross-section is
\begin{equation}
\sigma^I({\bf q})=\left(\frac{\gamma r_0}{2}\right)^2
	\langle M\rangle^2 \left|f(q)\right|^2 
	(1+|\widehat{\bf q}\cdot \widehat{\bf c}|^2).
\label{eq_cs1}
\end{equation}
For model II, let ${\bf x}={\bf a}\cos(\phi+\pi/4)+{\bf b}\sin(\phi+\pi/4)$,
i.e., magnetic moment pointing at an angle $\phi$ away from the (110) 
direction in the $a$-$b$ plane. 
In general, there can be eight magnetic twins. Assuming equal occupation among
the twins,
\[
\sigma^{II}({\bf q})=\frac{1}{2}\left(\frac{\gamma r_0}{2}\right)^2
	\langle M\rangle^2 \left|f(q)\right|^2 
	\left(1+|\widehat{\bf q}\cdot \widehat{\bf c}|^2
	\right).
\]
Thus $\sigma^{II}({\bf q})=\sigma^{I}({\bf q})/2$, and
model I and model II can not be distinguished in the diffraction. 
However, we prefer model I since 
a collinear magnetic
modulation (model II) usually squares up with lowering temperature, 
generating higher order harmonics\cite{jeff}. 
This is not what we have observed in CeRhIn$_5$.
Handedness of the magnetic
spiral ($\alpha=i$ or $-i$) can not be distinguished in this 
experiment.

\begin{figure}[bt]
\centerline{
\psfig{file=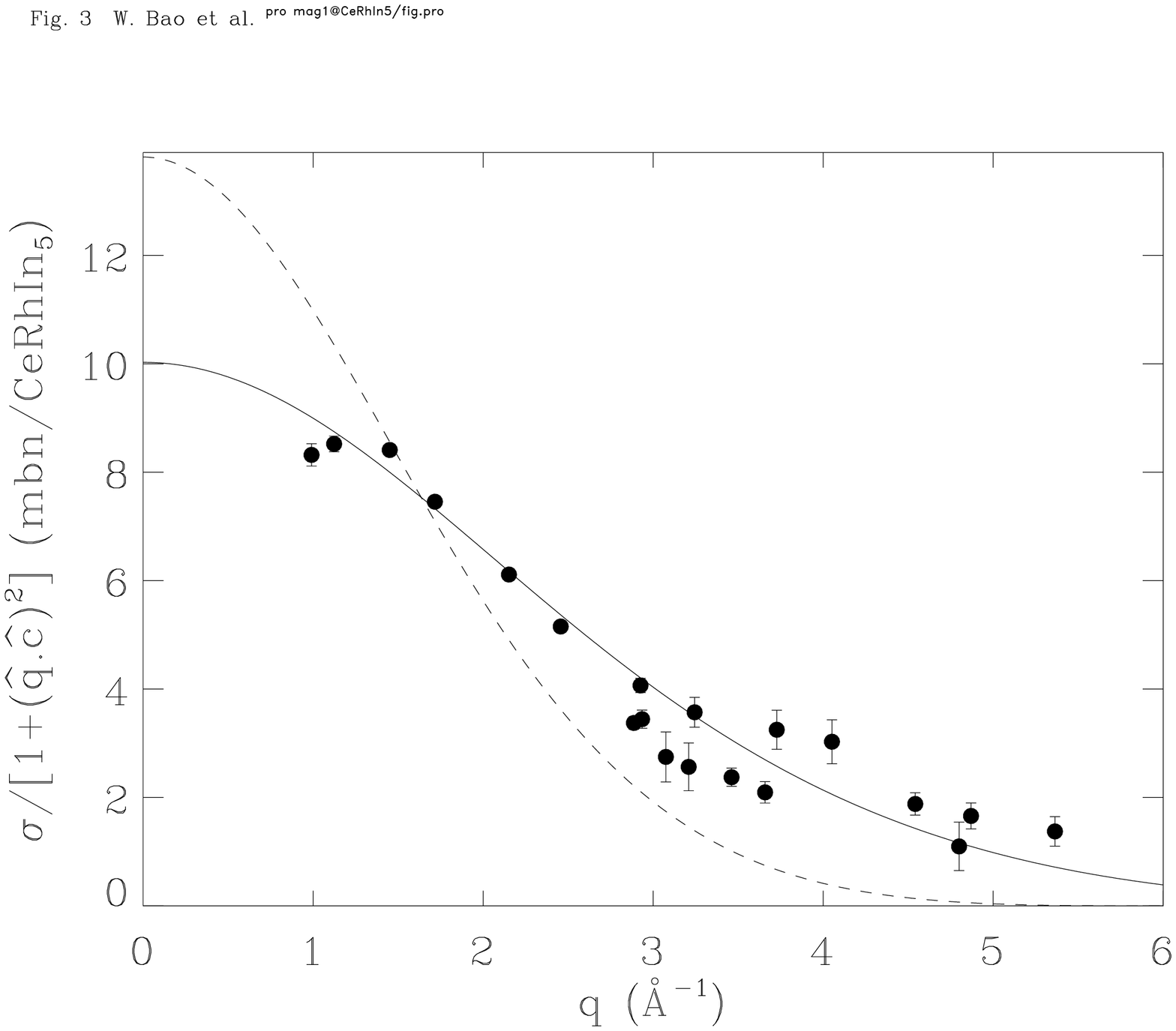,width=\columnwidth,angle=90,clip=}}
\caption{The $q$ dependence of magnetic cross-section, $\sigma$, divided by 
the polarization factor [refer to Eq.~(\ref{eq_cs1})]. The solid line is
Ce$^{3+}$ form factor\protect\cite{formf_ce}, 
and the dashed line Rh$^{+}$ form 
factor\protect\cite{formf_itc}.}
\label{formf}
\end{figure}

Fig.~\ref{formf} shows the quantity 
$\sigma({\bf q})/(1+|\widehat{\bf q}\cdot \widehat{\bf c}|^2)$ which
is proportional to atomic form factor $|f(q)|^2$ 
[refer to Eq.~(\ref{eq_cs1})].
The $\chi^2$ for best fit using Rh ion form factor (the dashed line)
is one order of magnitude
larger than that using the Ce one (solid line).
This indicates that the magnetic moments reside 
on the Ce ions rather than on the Rh ions.
The staggered moment at 1.4~K is $\langle M\rangle=0.374(5)\mu_B$ per Ce.

Within a CeIn$_3$ layer of CeRhIn$_5$, 
the local environment of a Ce ion is
similar to that in the cubic compound CeIn$_3$. It is interesting to note
that magnetic moments on these Ce share the same antiferromagnetic alignment.
The twist of magnetic moments along the $c$ axis in CeRhIn$_5$, 
approximately 107$^o$ per CeIn$_3$ layer, apparently is related
to the intervening RhIn$_2$ layer. 
From this diffraction work, we cannot definitely
tell whether this incommensurate twisting is due to competing
magnetic interactions in a localized moment model or a divergent
magnetic susceptibility at a nesting Fermi surface wave vector, 
i.e., a SDW. The
staggered moment of 0.374(5)$\mu_B$ per Ce in CeRhIn$_5$ is 
substantially reduced compared to that in its cubic counterpart CeIn$_3$.
This could be a sign of SDW order, but could also be caused
by a stronger Kondo effect or be a result of enhanced local
moment fluctuations
due to low dimensionality.
Revealing evidences may be obtained in an inelastic neutron
scattering experiment, as in the case of metallic V$_{2-y}$O$_3$\cite{bao93,bao96a}.

\begin{figure}[bt]
\centerline{
\psfig{file=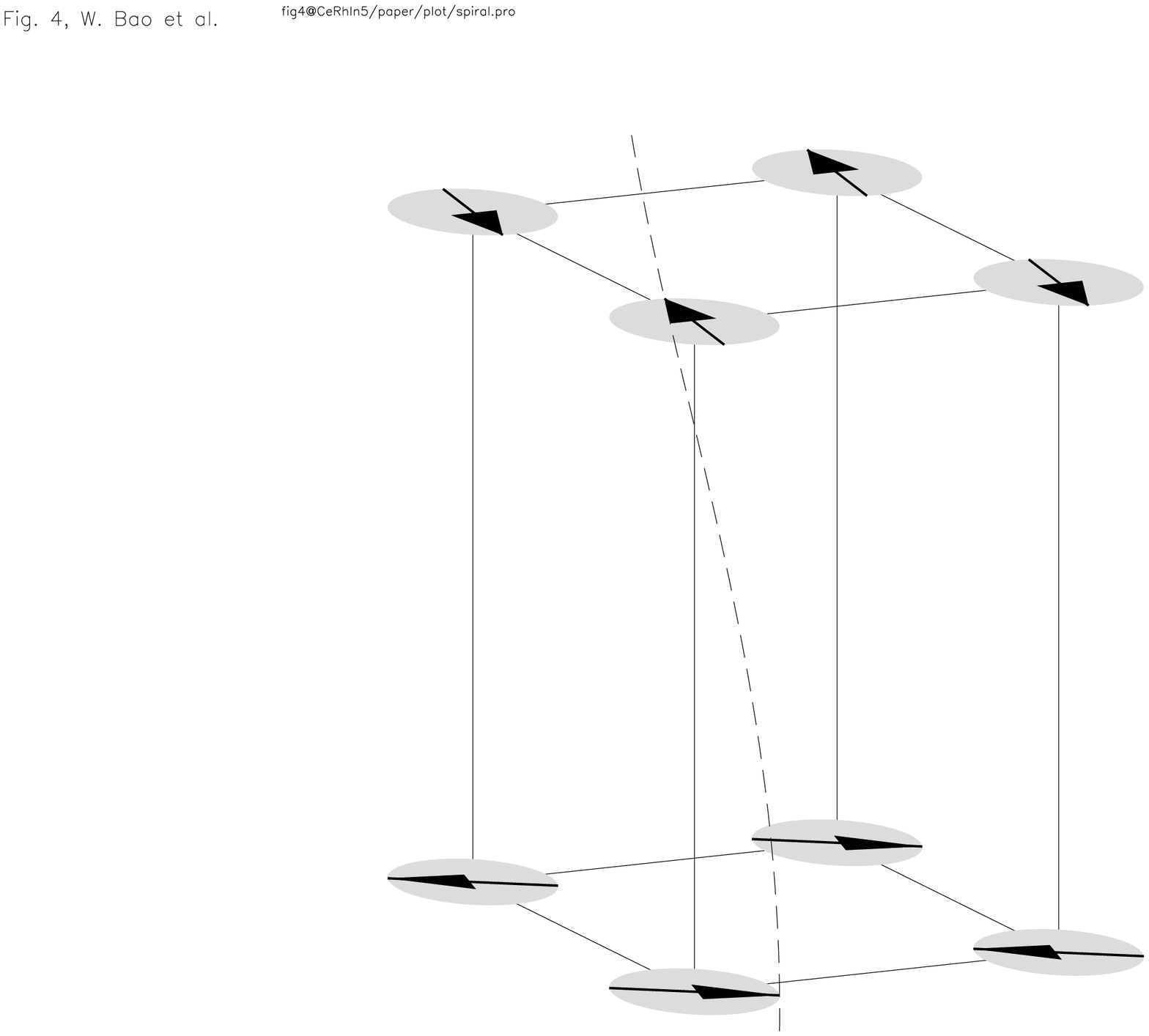,width=\columnwidth,angle=90,clip=}}
\caption{Magnetic structure of CeRhIn$_5$. Only Ce sites are shown in
the structural unit cell. The disk denotes the moment rotating plane.
The dashed line traces the spiral.}
\label{mstru}
\end{figure}

In conclusion, we find the incommensurate magnetic structure as 
depicted in Fig.~\ref{mstru} for CeRhIn$_5$. 
A magnetic moment of 0.374(5)$\mu_B$ resides on
the Ce ion at 1.4K and the $a$-$b$ plane is its easy plane. 
Within an $a$-$b$ plane, magnetic
moments form a simple nearest neighbor antiferromagnet on a square lattice, 
and they spiral transversely along the $c$ axis with an incommensurate
pitch $\delta$. The incommensurability, $\delta$, does not change
with temperature.

We thank G. Aeppli, S. M. Shapiro, and C. Broholm for valuable discussions.
Work at Los Alamos was performed under the auspices of the US Department
of Energy. ZF gratefully acknowledges NSF support at FSU. PGP acknowledges
FAPESP for partial support.

\end{document}